\documentclass[a4paper,fleqn,usenatbib,useAMS]{mnras}

\usepackage[T1]{fontenc}
\usepackage{ae,aecompl}

\usepackage{graphicx}	
\usepackage{amsmath}	
\usepackage{amssymb}	
\usepackage{bm}		
\usepackage{pdflscape}	

\usepackage{times}
\usepackage{txfonts}
\usepackage{color}

\newcommand{\fracsb}[2]{\left[\frac{#1}{#2}\right]}
\newcommand{\mean}[1]{\langle{#1}\rangle}

\title[Off-axis GRB afterglow light curves \& images]{Off-axis afterglow light curves and images from 2D hydrodynamic simulations of double-sided GRB jets in a stratified external medium}
  
\author[Granot, De Colle \& Ramirez-Ruiz]{
Jonathan Granot,$^{1,2}$\thanks{E-mail: granot@openu.ac.il (JG)}
Fabio De Colle$^{3}$
and Enrico Ramirez-Ruiz$^{4,5}$
\\
$^{1}$Department of Natural Sciences, The Open University of Israel, P.O Box 808, Ra'anana 43537, Israel\\
$^{2}$Department of Physics, The George Washington University, Washington, DC 20052, USA\\
$^{3}$Instituto de Ciencias Nucleares, Universidad Nacional Aut\'onoma de M\'exico, A. P. 70-543 04510 D. F. Mexico\\
$^{4}$TASC, Department of Astronomy \& Astrophysics, University of California, Santa Cruz, CA 95064, USA\\
$^{5}$Niels Bohr Institute, University of Copenhagen, Blegdamsvej 17, DK-2100 Copenhagen, Denmark
}

\date{Accepted 2018 September 05. Received 2018 September 05; in original form 2018 March 15}

\pubyear{2018}

\begin{document}
\label{firstpage}
\pagerange{\pageref{firstpage}--\pageref{lastpage}}
\maketitle

\begin{abstract}
Gamma-ray burst (GRB) jets are narrow, and thus typically point away from us. 
They are initially ultra-relativistic, causing their prompt $\gamma$-ray and early afterglow 
emission to be beamed away from us. However, as the jet gradually decelerates its beaming 
cone widens and eventually reaches our line of sight and the afterglow emission may be detected. 
Such orphan afterglows were not clearly detected so far. Nevertheless, they should be detected in 
upcoming optical or radio surveys, and it would be challenging to clearly distinguish between them 
and other types of transients. Therefore, we perform detailed, realistic calculations of the expected 
afterglow emission from GRB jets viewed at different angles from the jet's symmetry axis. 
The dynamics are calculated using 2D relativistic hydrodynamics simulations 
of jets propagating into different power-law external density profiles, 
$\rho_{\rm ext}\propto{}R^{-k}$ for $k=0,\,1,\,1.5,\,2$, ranging from a uniform 
ISM-like medium ($k=0$) to a stratified steady stellar-wind like profile ($k=2$). 
We calculate radio, optical and X-ray lightcurves, and the evolution of the radio afterglow image 
size, shape and flux centroid. This may help identify misaligned relativistic jets, whether 
initially ultra-relativistic and producing a GRB for observers within their beam, 
or (possibly intrinsically more common) moderately relativistic, in either 
(i) nearby supernovae Ib/c (some of which are associated with long duration GRBs), 
or (ii) in binary neutron star mergers, which may produce short duration GRBs,
and may also be detected in gravitational waves (e.g. GW$\,$170817/GRB$\,$170817A
with a weak prompt $\gamma$-ray emission may harbor an off-axis jet). 
\end{abstract}


\section{Introduction}

It has been realized early on \citep{Rhoads97} that the ultra-relativistic outflows that power 
GRBs are likely collimated into narrow jets, and therefore their prompt emission might be 
too dim to detect unless the jet is pointed towards us. However, during the afterglow phase
the jet decelerates by sweeping up the external medium and its emission is beamed into an 
increasing solid angle, and may become visible for observers at larger viewing angles $\theta_{\rm obs}$
from the jet's symmetry axis. 
Such an ``orphan afterglow'' without a detected prompt $\gamma$-ray
emission was not clearly detected yet and could potentially teach us a lot about the jet's angular structure
and degree of collimation \citep[e.g.][]{WL99,NPG02,TP02,Levinson02,HDL02,NP03,Rhoads03,GalYam06,ZWD07,Rossi08,vanEerten10b,Ghirlanda14,Lamb18}.

For convenience, most works assume a uniform conical jet with sharp edges at a half-opening angle 
$\theta_j$ with an initial value of $\theta_0$, often referrer to as a ``top hat jet''.
For such an initial jet angular structure, once the jet's Lorentz factor $\Gamma$ decreases below 
$1/\theta_0$  it comes into lateral causal contact and could start to significantly expand sideways, 
though the actual rate of lateral spreading is rather involved 
\citep[e.g.][]{Rhoads99,SPH99,Granot01,ZM09,Wygoda11,vanEerten12b,GP12}.
Moreover,  around the same time the jet's edge becomes visible for an observer along its symmetry 
axis ($\theta_{\rm obs}=0$). 
This leads to a steepening of the afterglow flux decay rate for such ``on-axis'' observers, 
known as a ``jet break'' \citep[e.g.][]{Rhoads97,Rhoads99,SPH99,PM99}. 
For  $0<\theta_{\rm obs}<\theta_0$ different parts of the jet's edge become 
visible at somewhat different times causing a smoother and somewhat later jet break \citep[e.g.][]{Granot01,vanEerten10b,DeColle12b,Ryan15}. 
For ``off-axis'' observers outside of the jet's initial aperture, $\theta_{\rm obs}>\theta_0$ 
(or $\Gamma_0(\theta_{\rm obs}-\theta_0)\gtrsim\;$a few, where $\Gamma_0$ is the initial Lorentz factor), 
the prompt GRB emission is strongly suppressed due to relativistic beaming, and is likely to be missed.

However, such a sharp outer edge for the jet is not very physical, and it is much more natural to expect
the initial energy per solid angle $\epsilon_0=dE_0/d\Omega$ (and possibly also $\Gamma_0$) in the jet to drop 
more gradually and smoothly outside of some jet core angle, $\theta_c$. Various different jet angular structures have 
been considered in the literature \citep[e.g.][]{MRW98,Rossi02,ZM02,KG03,Granot07,GR-R13}, where the 
most popular are a ``universal structured jet'' where $\epsilon_0(\theta>\theta_c)\propto\theta^{-2}$ 
and a Gaussian jet where $\epsilon_0\propto\exp(-\theta^2/2\theta_c^2)$, which can reproduce ``on-axis'' afterglow 
lightcurves that are broadly similar to observations and to those from a top hat jet \citep[some jet structures can 
be ruled out as they do not produce the observed ``on-axis'' afterglow lightcurves, e.g.][]{GK03,Granot05}. 
Because of the strong relativistic beaming during the prompt GRB emission (as $\Gamma_0\gtrsim100$ is 
typically required by compactness arguments) and the early afterglow, even a small amount of energy in outflow 
propagating towards an off-axis observer at the outer wings of the jet could dominate the observed flux over the 
strongly suppressed contribution from the much more energetic core of the jet. However, as the faster and more 
energetic parts of the jet near its core gradually decelerate as the jet sweeps up the external medium, they gradually
come into view as their beaming cone reaches the line of sight. If the jet's core contains the bulk of its energy 
(e.g. for an initially top hat jet or a Gaussian jet viewed from $\theta/\theta_c\gtrsim\;$a few) and $\epsilon_0$
rises steeply enough towards the jet's core, then the flux for an off-axis observer initially rises until the beaming 
cone of the jet's core reaches the line of sight,  and only then does the emission from the jet's core start to dominate 
the observed flux, which peaks around that time and starts to decay, approaching the lightcurve for an on-axis 
observer \citep[e.g.][]{Granot02,KG03,EG06}. 

Here we use numerical simulations of an initial top hat jet \citep{DeColle12a,DeColle12b}. 
Nonetheless, even such an initially top hat jet develops an egg-shaped bow shock structure on the 
dynamical time due to its interaction with the external medium \citep[e.g.][]{Granot01,ZM09}.
This makes it somewhat more realistic and interesting to compare with observations.
At early times the afterglow flux for an ``off-axis''  observer is dominated by emission from the slower
material at the sides of the jet, and it is relatively sensitive to the jet's initial angular structure.
However, once the beaming cone of the jet's core reaches the line of sight near the peak in the 
lightcurve it starts dominating the observed flux, which in turn becomes rather insensitive 
to the jet's initial angular structure outside of its core. Therefore, we expect that the results presented 
here should be broadly similar to those for other jet angular structures in which most of the jet's energy 
is contained within its narrow core \citep[see, e.g.,][]{DeColle18,GG18}. Moreover, such detailed 
properties of the afterglow lightcurves and image may help to more clearly distinguish between orphan 
GRB afterglows and other types of transients in upcoming surveys, which may otherwise be very challenging.

The main novelty of this work in calculating the off-axis afterglow emission for different viewing 
angles $\theta_{\rm obs}$ is (i) considering different external density profiles, namely
$\rho_{\rm ext}\propto{}R^{-k}$ for $k=0,\,1,\,1.5,\,2$, and (ii) calculating in addition to the off-axis
afterglow lightcurves also the corresponding afterglow images, and in particular the flux centroid motion
and the evolution the image size and shape, which may be more readily compared to observations when 
the image is marginally resolved. Such relatively realistic and detailed calculations may be very useful 
for identifying orphan GRB afterglows within the zoo of different transients expected in upcoming 
surveys (also in the optical, e.g. LSST).

In \S~\ref{sec:LC} we present radio, optical and X-ray afterglow lightcurves for a wide range of viewing angles 
$\theta_{\rm obs}$ for a jet propagating into a power law external density $\rho_{\rm ext}\propto{}R^{-k}$ 
ranging from a uniform  ISM-like medium ($k=0$) to a profile expected for a steady stellar wind ($k=2$). 
In \S~\ref{sec:images} we calculate the corresponding afterglow images in the radio and show the evolution 
of the image size, shape and  flux centroid. There are two main motivations behind this. 
First,  this may help identify misaligned relativistic jets in  nearby supernovae Ib/c 
\citep{GL03,GR-R04,SFW04,GR-RL05,R-R05a,Bietenholz10,Bietenholz14,XNW11,Sobacchi17} 
that are either (i) initially ultra-relativistic jets that produce a long GRB whose prompt $\gamma$-ray emission 
is strongly beamed away from us, or (ii) initially mildly relativistic jets,  that may be more numerous.
Second, in order to help infer the presence of a relativistic jet in compact binary mergers involving one or two 
neutron stars (NS-NS or NS-BH), and constrain our viewing angle and the jet's angular structure
\citep{Rezzolla11,M-B14,Nagakura14,Duffell15,Ruiz16,M-B17,Lazzati17a,Lazzati17b,LK17}. 
This is naturally also motivated by the recent binary neutron star merger GW$\,$170827/GRB$\,$170817A
that was detected in gravitational waves and had a weak prompt $\gamma$-ray emission and still
shows a rising afterglow lightcurve from radio to X-rays 
\citep[e.g.][]{Abbott17a,Abbott17b,Abbott17c,Goldstein17,Haggard17,Margutti17,Hallinan17,Drout17,LK18,Ruan18,Margutti18,Lyman18,Mooley18,Lazzati18,Lazzati18,NP18}.
In  \S~\ref{sec:scaling} we discuss the scaling of our results with the model parameters, 
and how our results may help break degeneracies between the model parameters.
Our conclusion are discussed in \S~\ref{sec:dis}

\section{Off-Axis Afterglow Lightcurves For  a Jet in a Power-Law External Density Profile}
\label{sec:LC}

\begin{figure*}
\begin{center}
\includegraphics[width=1.02\columnwidth,height=10.8cm]{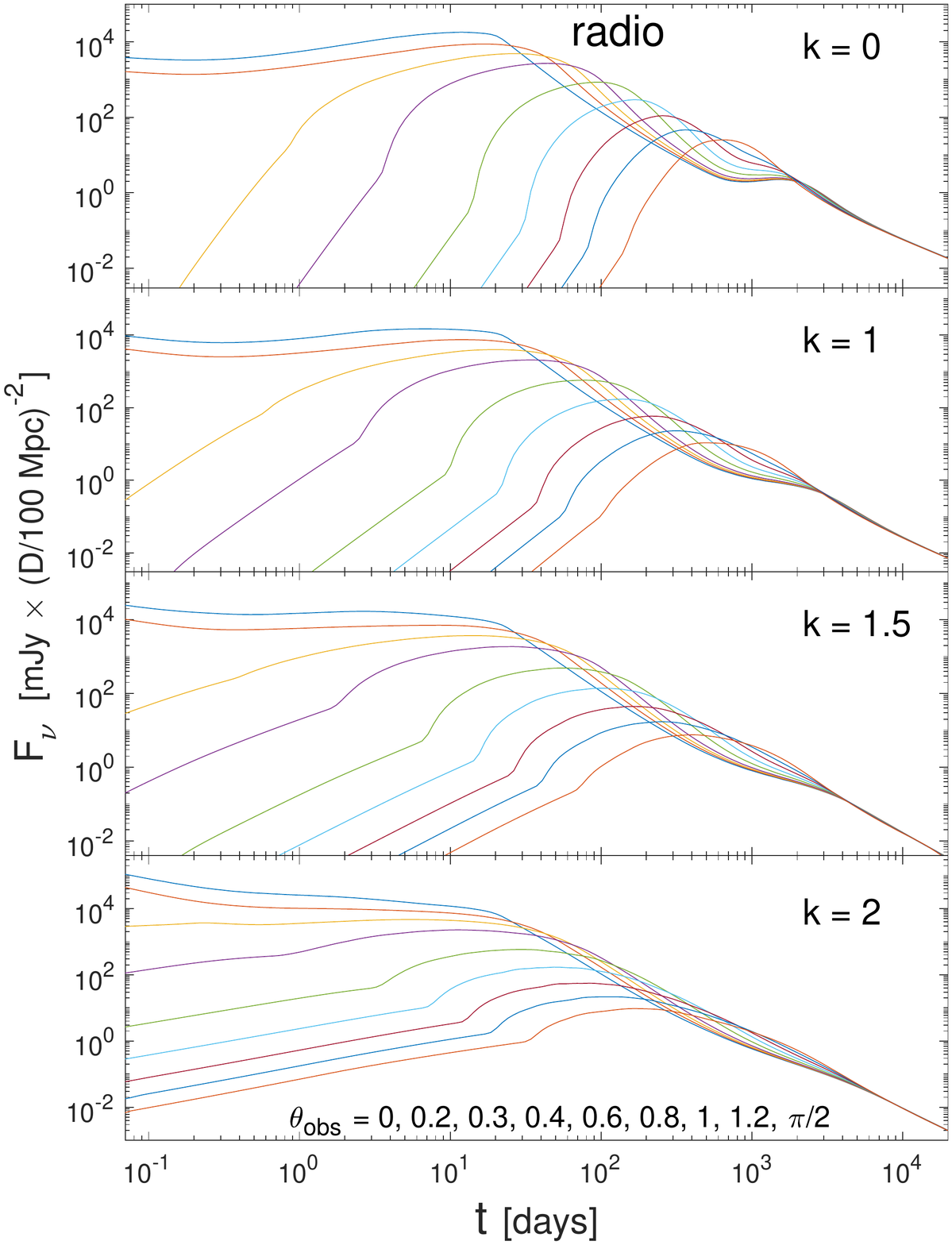}
\hspace{0.30cm}
\includegraphics[width=1.02\columnwidth,height=10.8cm]{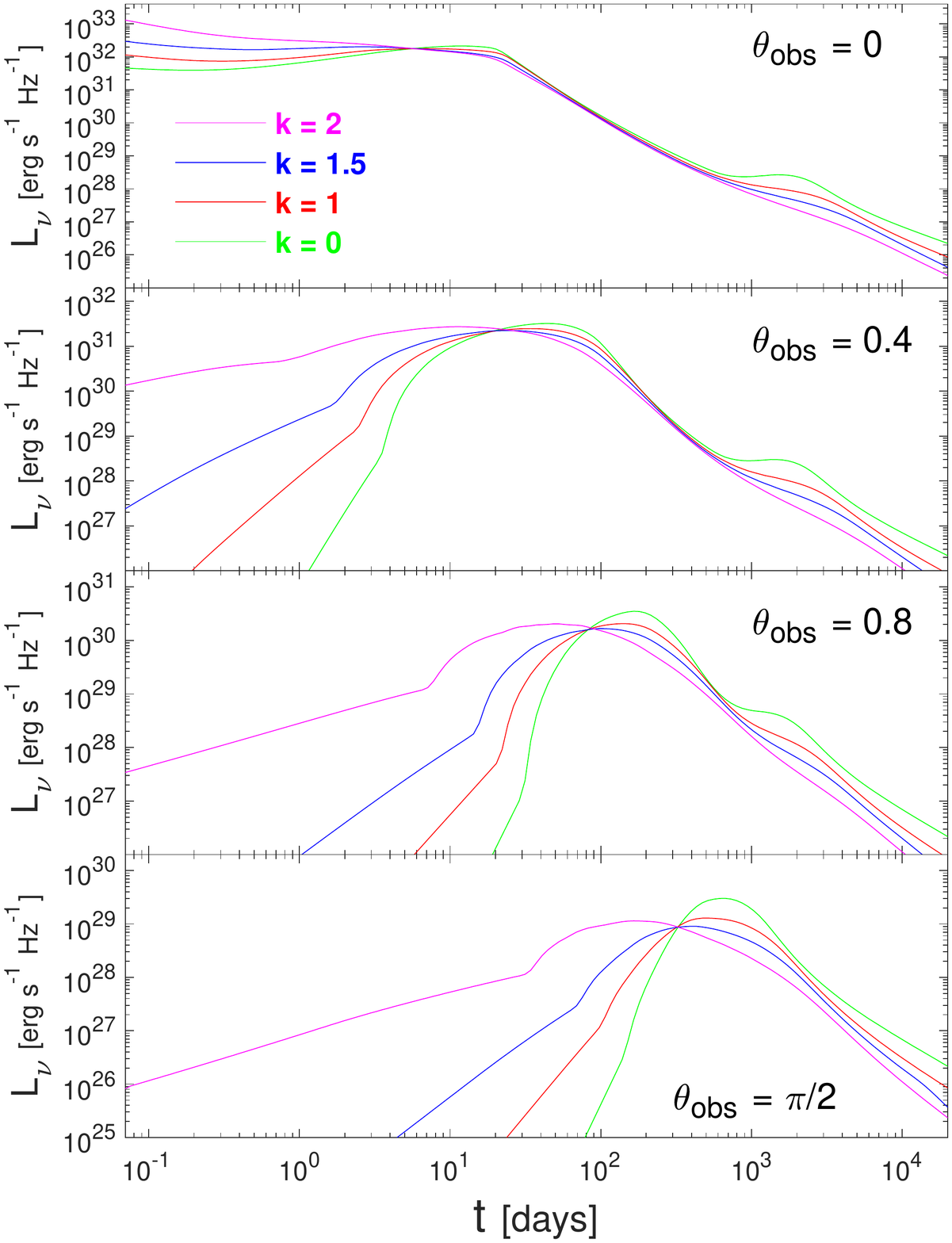}
\caption{Radio light curves ($\nu=8.46\;$GHz) for hydrodynamic 
simulations of an initially conical jet (see text for details).  
{\bf Left:} each panes corresponds to a different value of the external density power-law index,
$k=0,\,1,\,1.5,\,2$ where $\rho_{\rm ext}=AR^{-k}$, and shows
lightcurves for different viewing angles, $\theta_{\rm obs}=0,\,0.2,\,0.3,\,0.4,\,0.6,\,0.8,\,1,\,1.2,\,\pi/2$. 
{\bf Right:} each panel corresponds to a different viewing angle, $\theta_{\rm obs}=0,\,0.4,\,0.8,\,\pi/2$
from top to bottom, and shows lightcurves for different values of the external density power-law index, 
$k=0,\,1,\,1.5,\,2$.}
\label{fig:off-axis_radio}
\end{center}
\end{figure*}

For the calculations presented here we use 2D hydrodynamic simulations from 
\citet{DeColle12b}, based on the special relativistic hydrodynamics code \emph{Mezcal}, 
and a complimentary code for calculating the radiation by post-processing the results 
of the numerical simulations \citet{DeColle12a}. The initial conditions for the GRB jet 
were a conical wedge of half-opening angle $\theta_0=0.2\;$rad, taken out of the 
spherical self-similar \citet{BM76} solution. The simulation starts when the Lorentz 
factor of the material just behind the shock is $\Gamma=20$.
The calculation of the synchrotron radiation is supplemented by
adding the contribution from a \citet{BM76} conical wedge at earlier times, 
corresponding to $20\leq\Gamma\leq500$ (which causes an artificially sharp 
transition in the light curve between the two at a rather early time). The simulation was for 
an isotropic equivalent kinetic energy of $E_{\rm k,iso}=E_{53}10^{53}\;$erg with $E_{53}=1$, 
corresponding to a true energy of $E_{\rm jet}=(1-\cos\theta_0)E_{\rm k,iso}\approx2\times10^{51}\;$erg 
for a double-sided jet. 

We consider synchrotron emission from relativistic electrons that are accelerated at the 
afterglow shock and radiate as the gyrate in the magnetic field within the shocked region. 
The microphysics processes responsible for magnetic field amplification and particle 
acceleration are parameterized here by assuming that the magnetic field everywhere in the 
shocked region holds a fraction $\epsilon_B = 0.1$ of the local internal energy density in
the flow, while the non-thermal electrons just behind the shock hold a fraction $\epsilon_e = 0.1$
of the internal energy, and have a power-law energy distribution, $N(\gamma_e)\propto\gamma_e^{-p}$
for $\gamma_e>\gamma_m$ with $p = 2.5$. For more details on the exact form of the spectral emissivity
that is used and the calculation of the lightcurves and images see \citet{DeColle12a,DeColle12b}.

The external density was taken to be a power law with radius,  $\rho_{\rm ext}=A_k r^{-k}$. 
We have made calculations for $k=0,\,1,\,1.5,\,2$, that cover the expected density profiles both for 
short GRBs, where a uniform ISM ($k=0$) is expected  (in particular for compact binary merger progenitors), 
and for long GRBs whose immediate circumburst medium is shaped by the stellar wind of their massive star
progenitors \citep[e.g.,][]{CL00,R-R01}, where $k=2$ corresponds to a steady wind, while variations in 
the wind's velocity and/or mass loss rate near the end of the massive star's life could lead to other values 
of $k$ \citep[e.g.,][]{G-S96,CLF04,R-R05b,vanMarle06}. For example, $k=1.4\pm0.2$ was inferred 
for the afterglow of the long and very bright GRB$\;$130427A \citep{Kouveliotou13}. 
The density normalization $A_k$ for the case $k=0$ (a uniform medium) was set to be 
$A_0=\rho_0=n_{\rm ext}m_p=1.67\times10^{-24}\;{\rm g\;cm^{-3}}$  corresponding to 
$n_0=n_{\rm ext}/(1\,{\rm cm^{-3}}) = 1$, while for other $k$-values it was set such that the density 
would be the same at the jet break radius \citep[corresponding approximately to the Sedov radius for a 
spherical flow with the same true energy; for details see][]{DeColle12b}. This corresponds to 
$A_*\equiv A/(5\times 10^{11}\;{\rm gr\;cm^{-1}})=1.65$ for $k=2$.

\begin{figure*}
\begin{minipage}{1.0\textwidth}
\includegraphics[width=0.48\columnwidth,height=10.8cm]{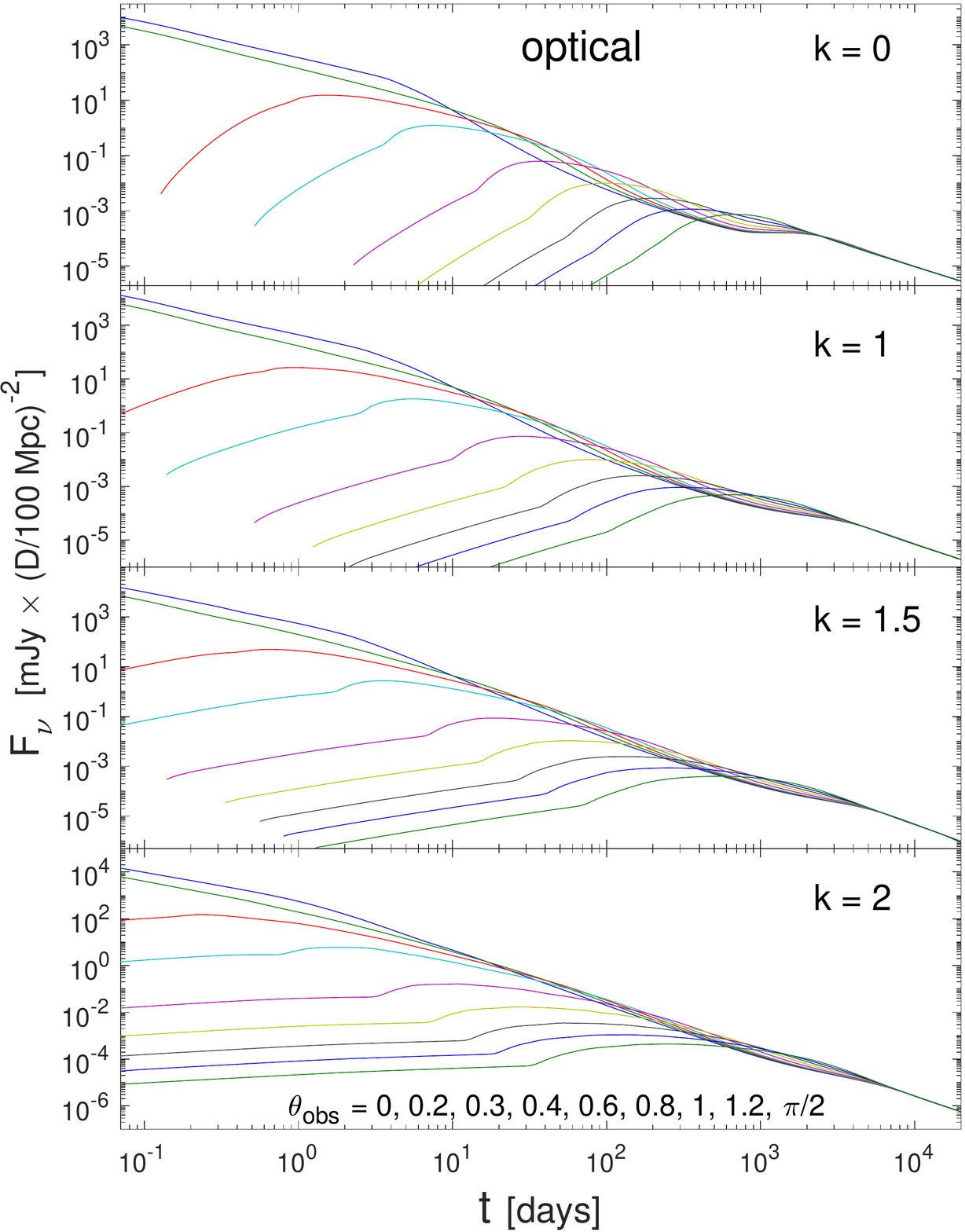}
\hspace{0.65cm}
\includegraphics[width=0.48\columnwidth,height=10.8cm]{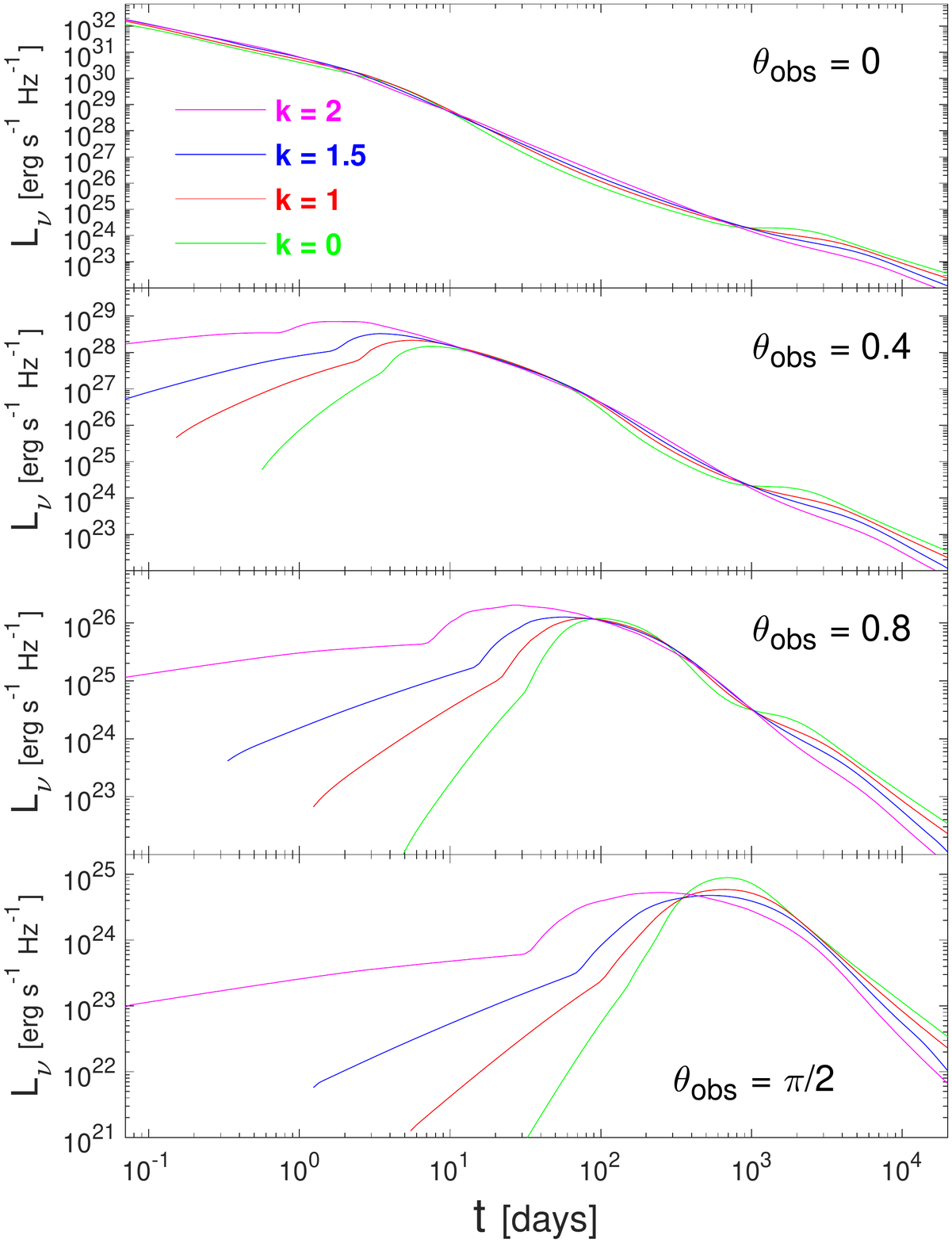}
\vspace{-0.3cm}
\caption{\label{fig:off-axis_opt} Optical  ({\bf left}; $\nu=4.56\,\times10^{14}\,$Hz, R-band) 
afterglow light curves, in the same format as Fig.$\,$\ref{fig:off-axis_radio}.}
\vspace{1.0cm}
\end{minipage}
\begin{minipage}{1.0\textwidth}
\includegraphics[width=0.48\columnwidth,height=10.8cm]{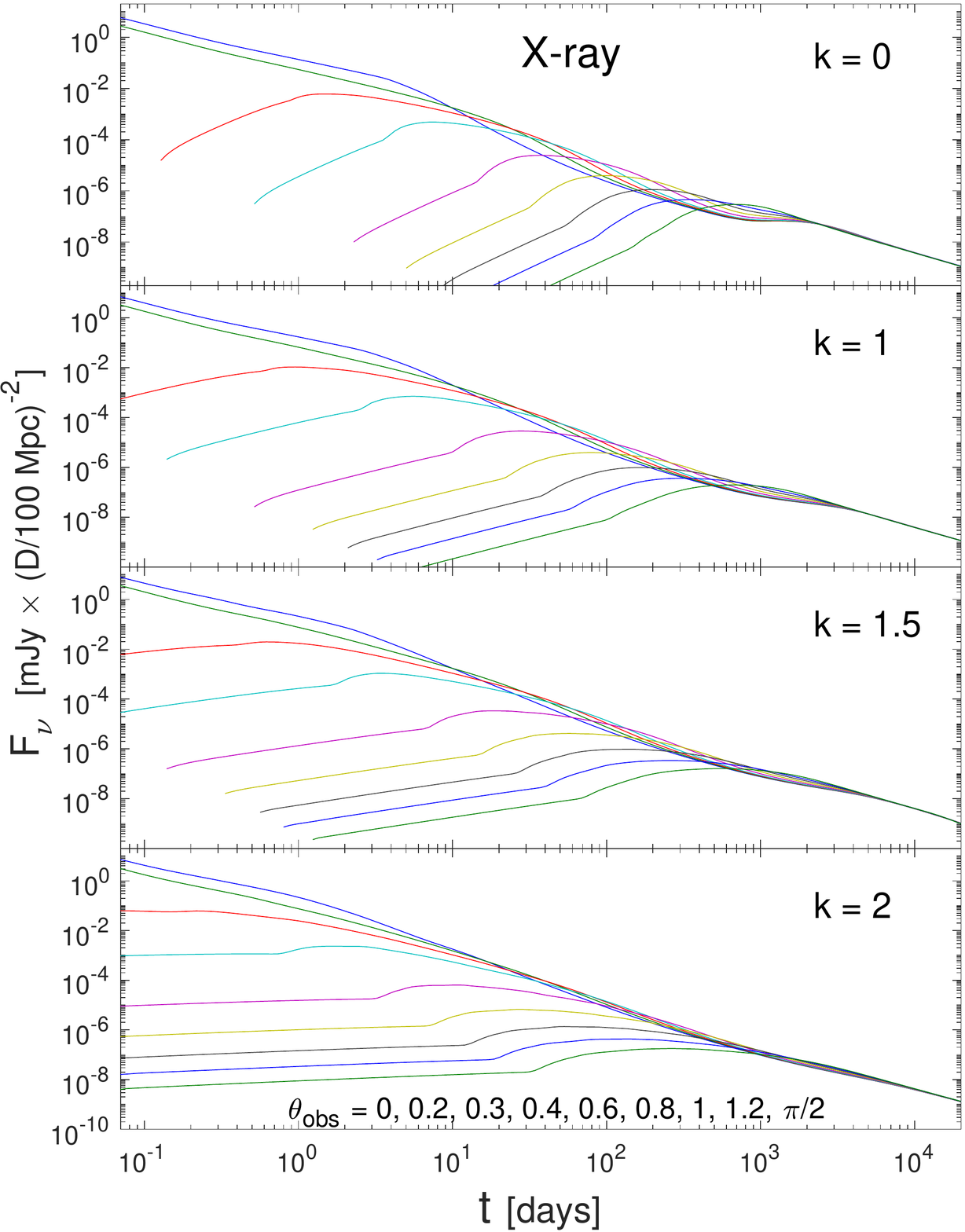}
\hspace{0.65cm}
\includegraphics[width=0.48\columnwidth,height=10.8cm]{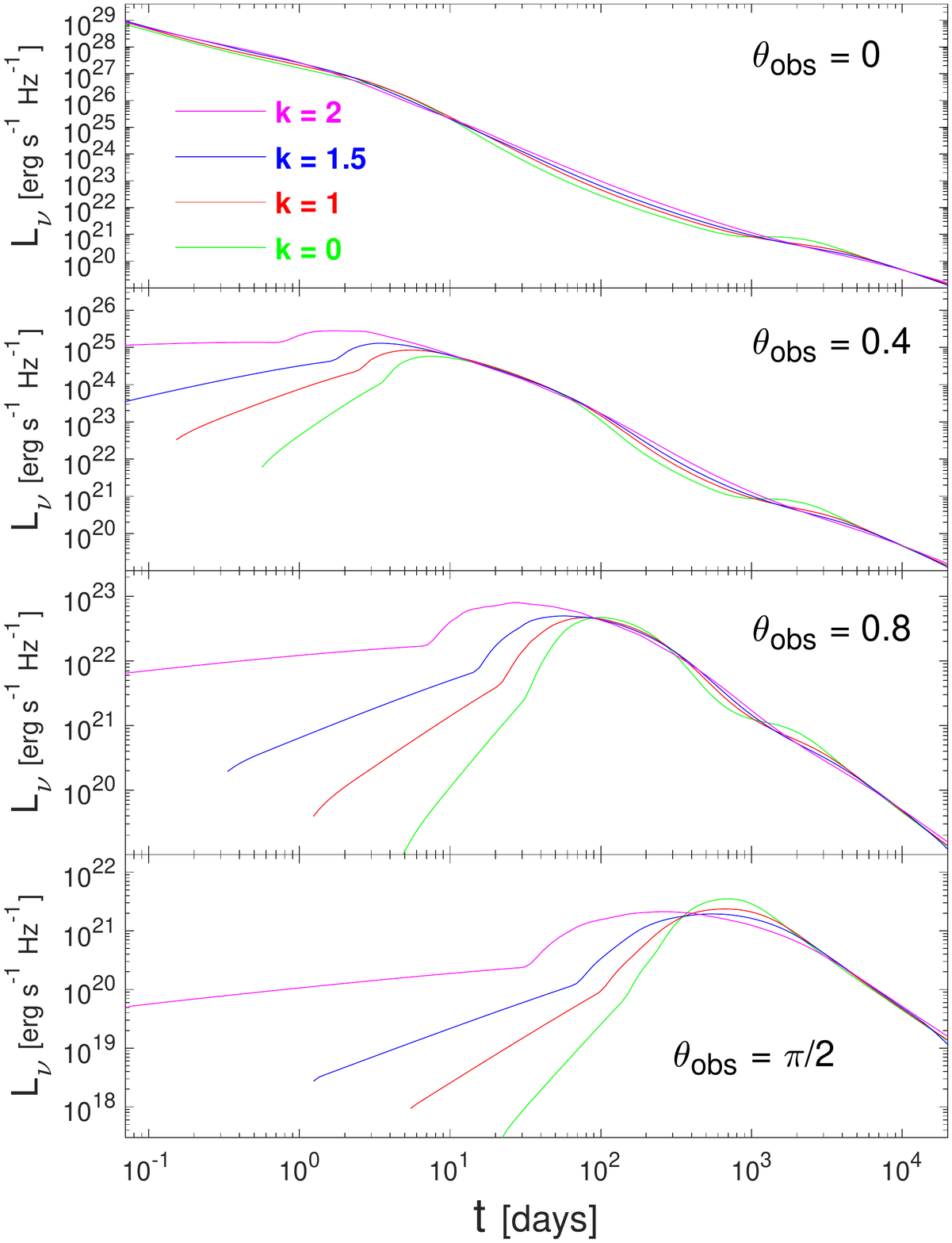}
\vspace{-0.3cm}
\caption{\label{fig:off-axis_Xray}X-ray ({\bf right}; $h\nu=1\,$keV, $\nu = 2.42\times 10^{17}\;$Hz) 
afterglow light curves, in the same format as Fig.$\,$\ref{fig:off-axis_radio}.}
\end{minipage}
\end{figure*}

The radio lightcurves for a wide range of viewing angles $\theta_{\rm obs}$ are shown in Fig.~\ref{fig:off-axis_radio}. Self-absorption is not included (but it is unimportant in the displayed times and frequency).
The left panels show lighcurves for a fixed $k$ and different $\theta_{\rm obs}$,
while the right panels show lighcurves for a fixed $\theta_{\rm obs}$ and different $k$.
Figs.~\ref{fig:off-axis_opt} and \ref{fig:off-axis_Xray} show  the afterglow lightcurves in the optical and X-ray, 
respectively, in the same format as Fig.~\ref{fig:off-axis_radio}.
The on-axis ($\theta_{\rm obs}=0$) jet break time is around $t_j\approx4-5\;$days, 
as can clearly be seen in the on-axis optical and X-ray lightcurves. In the radio the flux still keeps
gradually rising after $t_j$ until the passage of the typical synchrotron frequency $\nu_m$ through 
the observed frequency range, after which the flux decays similarly to the optical \citep{Granot01}.
 
For off-axis observers ($\theta_{\rm obs}>\theta_0$), the larger the external density power-law 
index $k$ the shallower the rise to the peak of the lightcurve, and the flatter and wider the peak. 
This more gradual evolution arises since for larger $k$ it takes a longer time to sweep up the same 
amount of external mass (for a spherical flow the accumulated swept up mass scales as $R^{3-k}$) 
that is needed in order for the jet to decelerate down to the same Lorentz factor with the same 
associated degree of relativistic beaming of the emitted radiation. 
For the same reason,  the bump in the afterglow lightcurve when the counter-jet becomes visible 
is much less pronounced for larger $k$-values, and it is very hard to clearly see it for $k=2$. 
This was shown for  an on-axis observer ($\theta_{\rm obs}=0$) in \citet{DeColle12b}, 
and here we find that this indeed persists for all $\theta_{\rm obs}<\pi/2$
(for $\theta_{\rm obs}=\pi/2$ the peak of the emission from the two sides of the jet exactly
coincides, as in this case they are both viewed from the same angle, resulting in a single peak).

The effect on the lightcurves of varying $k$ becomes smaller in the X-ray compared to the optical
or radio, since above the cooling break frequency, $\nu_c$, the observed flux density $F_\nu$ 
becomes much less sensitive to the external density $\rho_{\rm ext}$. We are in the slow cooling regime
($\nu_m<\nu_c$) so this corresponds to the power-law segment PLS H of the afterglow synchrotron spectrum
where $F_\nu\propto\nu^{-p/2}$ \citep{SPN98,GS02}, and for a relativistic self-similar flow \citep{BM76}
$F_\nu^{(H)}$ is independent of the external density. Once the flow becomes Newtonian and approaches 
the spherical Sedov-Taylor solution, there is some dependence of  $F_\nu^{(H)}$ on $\rho_{\rm ext}$.
However, it is a rather weak dependence, $F_\nu^{(H)}\propto\nu^{-p/2}\rho_{\rm ext}^{-(p-2)/4}$ 
at a given observed time $t$, with an exponent of $-1/8$ for $p=2.5$ or $-1/20$ for $p=2.2$.
For comparison, in PLS G where $\nu_m<\nu<\nu_c$,  
$F_\nu^{(G)}\propto\nu^{(1-p)/2}\rho_{\rm ext}^{1/2}$ for the relativistic spherical phase,
and  $F_\nu^{(G)}\propto\nu^{(1-p)/2}\rho_{\rm ext}^{(19-5p)/20}$ for the Newtonian spherical (Sedov-Taylor),
corresponding to an exponent of $0.325$ for $p=2.5$ or $0.4$ for $p=2.2$.
For this reason, a wind termination shock where the density switches from $k=2$ up to the 
termination shock radius and then becomes uniform ($k=0$, with a factor of 4 jump in the density at the shock)
is hardly seen in PLS H,  but in PLS G it is manifested as a flattening of the lightcurve by a factor of $t^{1/2}$
\citep{NG07}, which may partly mimic the effect of energy injection.

The bump or flattening in the lightcurve when the counter jet becomes visible can still be seen in the X-ray
for $k=0$ (and is much harder to see for larger $k$-values, similar to the optical or radio), since it arises 
from relativistic beaming, which is present in all spectral regimes as it is a dynamical effect. 

\begin{figure*}
\begin{center}
\includegraphics[width=0.95\textwidth]{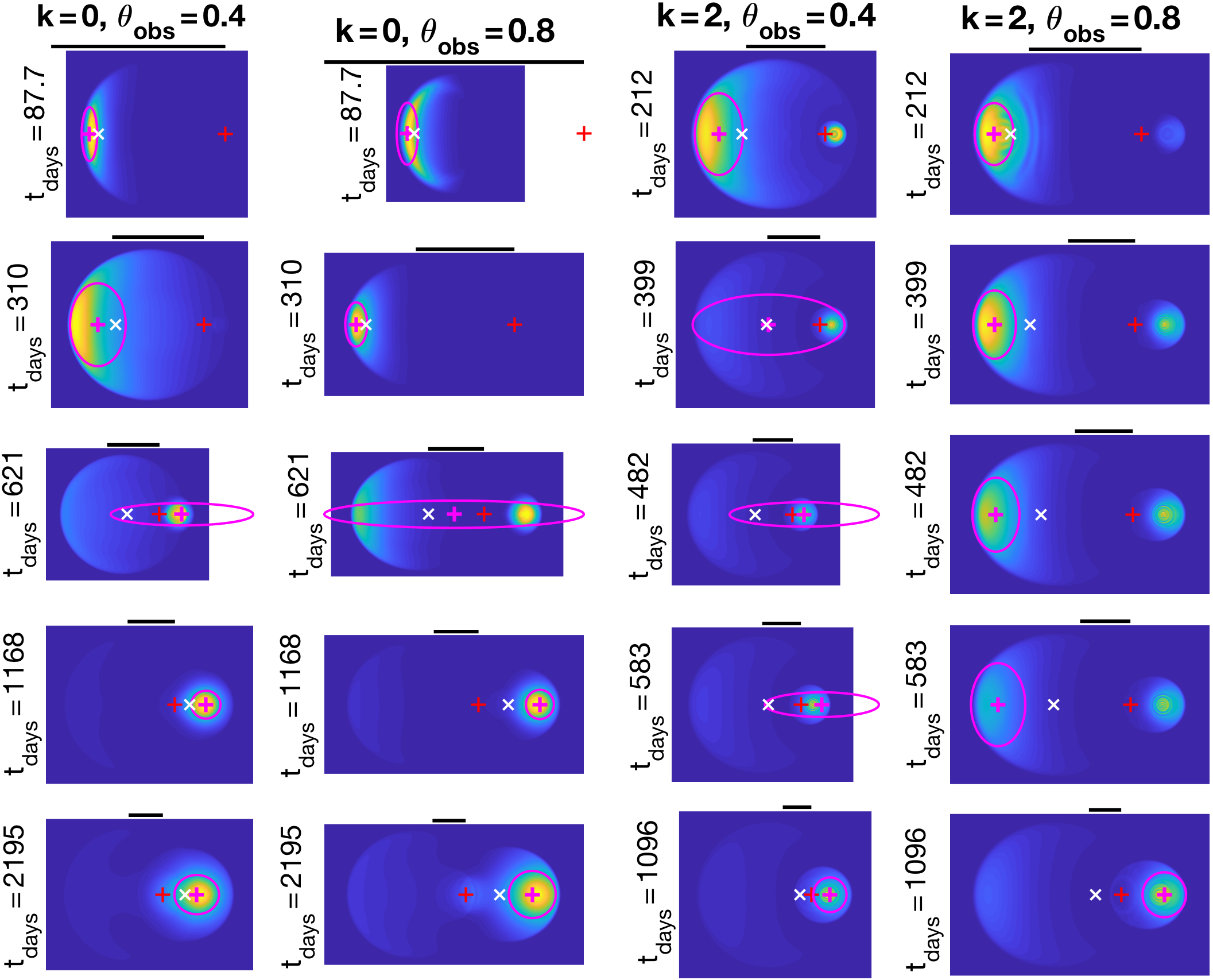}
\caption{Radio images (in PLS G, $\nu_m<\nu<\nu_c$) for $k=0$ (columns 1 and 2, from the left) 
and $k=2$ (columns 3 and 4),
for $\theta_{\rm obs}=0.4$ (columns 1 and 3) and $\theta_{\rm obs}=0.8$ (columns 2 and 4),
for five different observed times (rows 1 to 5). The thin red plus sign is the location of the central source. 
The thin white X sign is the location of the flux centroid. 
The magenta ellipse whose center is the thick magenta plus sign is the best fit
elliptical Gaussian to the image. The thick black line above each panel is a yardstick of length $10^{18}\;$cm.
Under a rescaling of the energy and $A_k$ by factors of $\zeta=E_{53}$ and $a$, respectively, the yardstick's
length becomes $\alpha\times10^{18}\;$cm where $\alpha = (E_{53}/a)^{1/(3-k)}$, and the observed time of
the image becomes $\alpha$ times that indicated in the figure (see \S\,\ref{sec:scaling} for details).}
\label{fig:images_k02}
\end{center}
\end{figure*}

\section{The Afterglow Image Size, Shape and Flux Centroid Evolution}
\label{sec:images}

\begin{figure}
\begin{center}
\includegraphics[width=0.60\columnwidth]{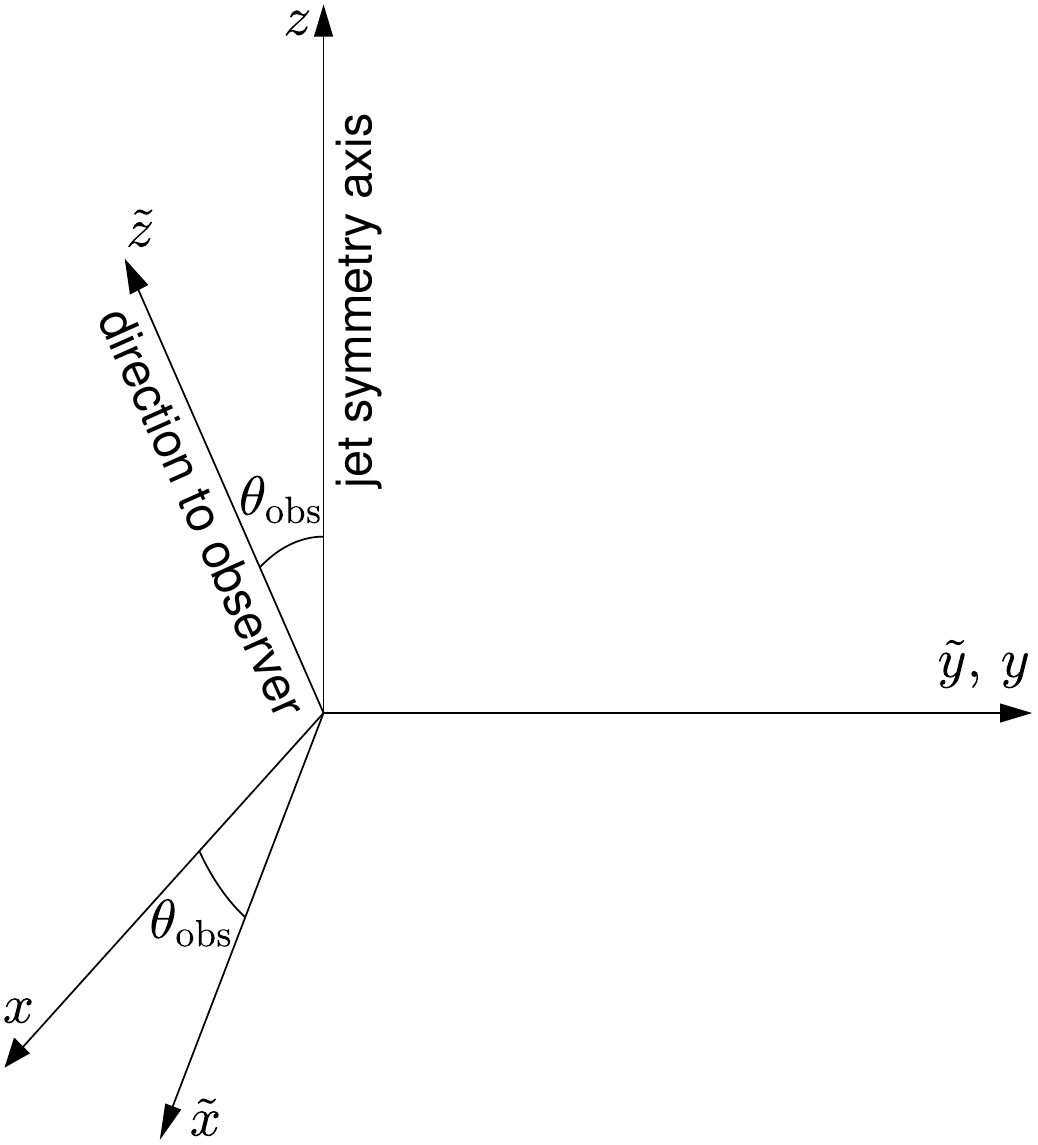}
\caption{A diagram of the coordinates we use. The $z$-axis is the jet's symmetry axis, while the 
$\tilde{z}$-axis points to the observer and is in the $x$-$z$ plane at an angle of $\theta_{\rm obs}$ 
from the $z$-axis. The $y$ and $\tilde{y}$ axes coincide. The afterglow image is in the plane of the sky, 
i.e. in the $\tilde{x}$-$\tilde{y}$ plane.}
\label{fig:coord}
\end{center}
\end{figure}

The afterglow image has so far been calculated mainly for a spherical flow
\citep[e.g.,][]{Waxman97,Sari98,PM98,GPS99a,GPS99b,Granot08,Morsony09,vanEerten10a}.
A few works have considered the afterglow images from a GRB jet \citep[e.g.,][]{IN01,Salmonson03,GG18}
or the flux centroid motion \citep[e.g.][]{Sari99,IN01,GL03}, but have used a simple analytic model 
for the jet dynamics. Here we consider the afterglow images from hydrodynamic simulations of the 
GRB jet in different external density profiles.

\begin{figure*}
\begin{center}
\includegraphics[width=1.02\columnwidth]{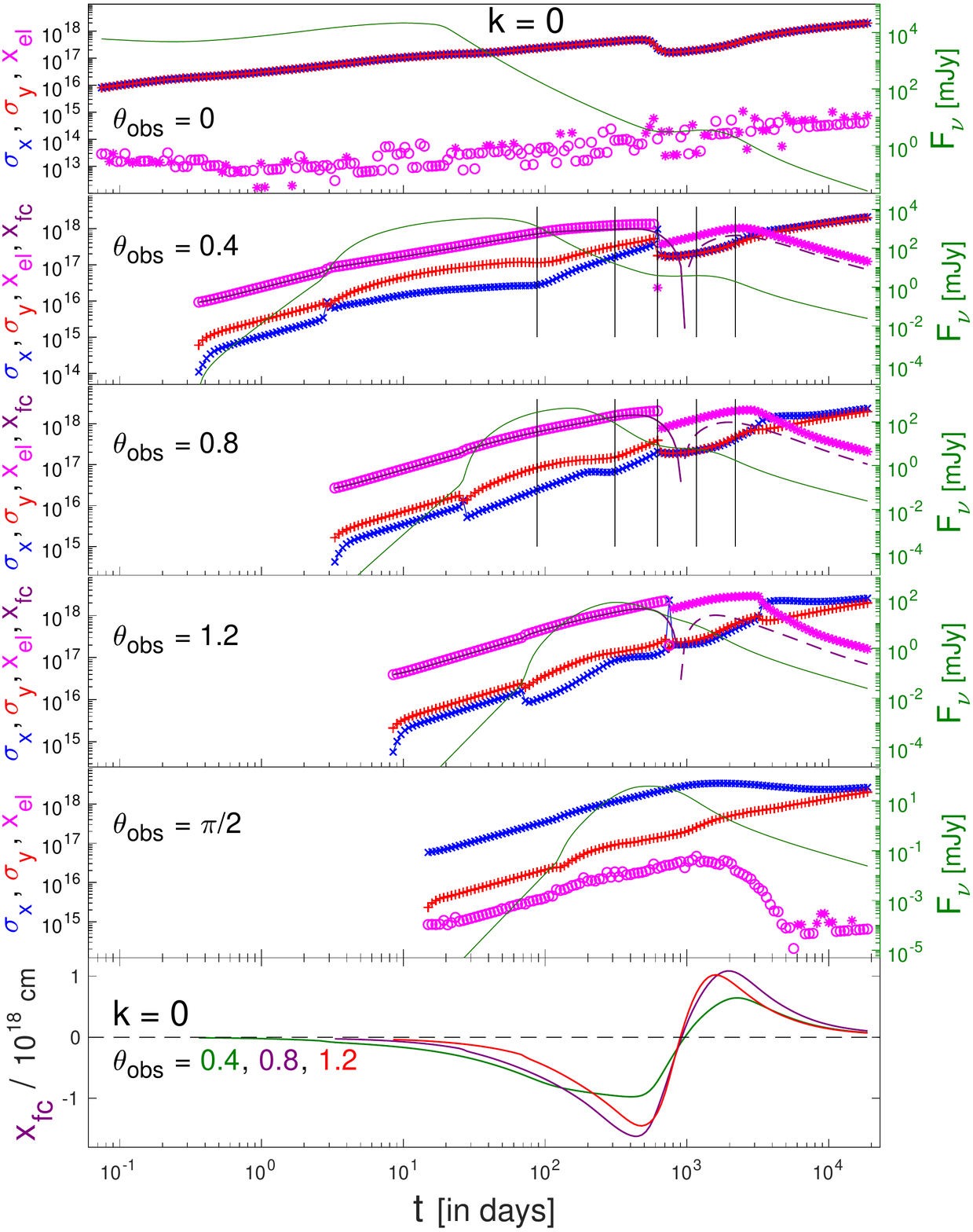}
\hspace{0.30cm}
\includegraphics[width=1.02\columnwidth]{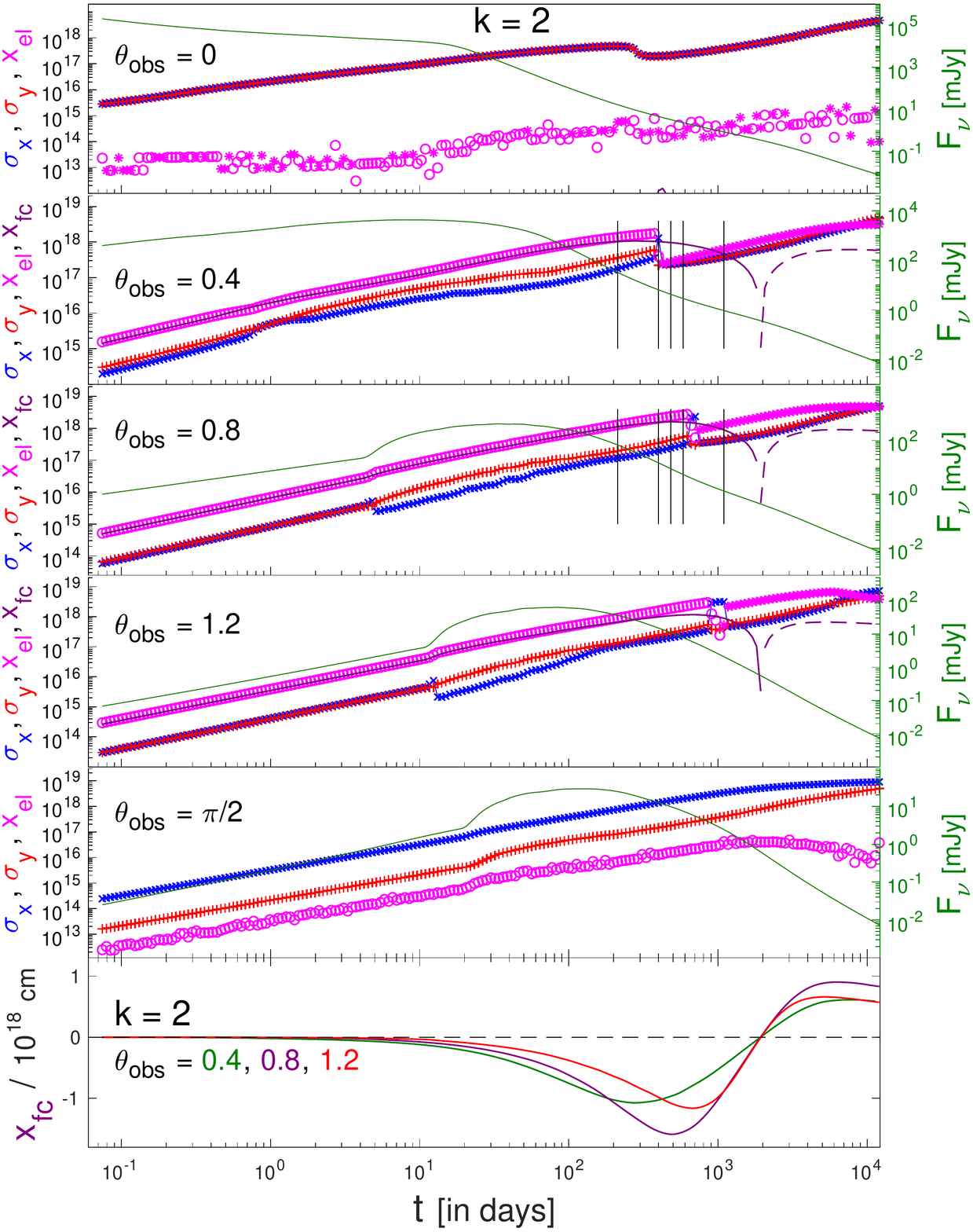}
\caption{Properties of the radio image observed from different viewing angles 
($\theta_{\rm obs}=0,\,0.4,\,0.8,\,1.2,\,\pi/2$.), for $k=0$ 
({\bf left}) and $k=2$ ({\bf right}). The flux density ({\it right $y$-axis}) is for $\nu = 8.46\;$GHz 
and a distance of $100\;$Mpc, while the normalized image ($I_\nu/\langle I_\nu\rangle$) 
holds for any frequency in PLS G, where $F_\nu\propto\nu^{(1-p)/2}$.
An elliptical Gaussian to the image is shown in terms of its best fit  parameters: the location 
of the ellipse's center, $\tilde{x}_{\rm el}$ (in {\it magenta}; $\tilde{x}_{\rm el}>0$ values are denoted 
by an asterisk, while $|\tilde{x}_{\rm el}|$ for $\tilde{x}_{\rm el}<0$ values is denoted by a circle), 
and its semi-axes $\sigma_x$ ({it blue} $x$'s) and $\sigma_y$ ({\it red} +'s). 
The symmetry of the problem implies $\tilde{y}_{\rm fc}=\tilde{y}_{\rm el}=0$.
Also shown as a useful reference are the radio lightcurves (in {\it dark green} solid lines, 
using the right $y$-axis). The vertical thin solid black lines in the second ($\theta_{\rm obs}=0.4$) 
and third ($\theta_{\rm obs}=0.8$) panels indicate the observed times for which the images 
are shown in Fig.~\ref{fig:images_k02}. The location of the flux centroid, $\tilde{x}_{\rm fc}$ (measured in cm; 
see Fig.~\ref{fig:coord}), is shown in {\it deep purple} ($\tilde{x}_{\rm fc}>0$ values are shown by the 
solid line while $|\tilde{x}_{\rm fc}|$ for $\tilde{x}_{\rm fc}<0$ values is shown by the dashed line). 
In the bottom panels $\tilde{x}_{\rm fc}$ is shown with a linear $y$-axis
(for $\theta_{\rm obs}=0,\,\pi/2$ one has $\tilde{x}_{\rm fc}=0$ due to symmetry). }
\label{fig:images_size_shape_k0}
\end{center}
\end{figure*}

Fig.~\ref{fig:images_k02} shows examples of images for two different
viewing angles ($\theta_{\rm obs}=0.4,\,0.8$) , and two different external density profiles: 
a uniform density ($k=0$) and a (steady) wind-like stratified medium ($k=2$). 
The coordinates we use for displaying the afterglow image on the plane of the sky
are shown in Fig.~\ref{fig:coord}, and follow section 3.2 of \citet{DeColle12a}.
The images are for PLS G ($\nu_m<\nu<\nu_c$), which typically applies to radio frequencies 
at reasonably late times in which the image may be resolved under favorable conditions. Note that 
within each PLS the normalized image (i.e. the specific intensity normalized by its mean value 
over the entire image, $I_\nu/\langle I_\nu\rangle$) is independent of the observing frequency 
\citep[e.g.][]{Sari98,GPS99a,GL01}.
The image is symmetric to reflection on the plane containing the jet symmetry axis ($z$-axis)
and the direction to the observer ($\tilde{z}$-axis), i.e. $\tilde{y}\to-\tilde{y}$.
The images in  Fig.~\ref{fig:images_k02} are shown at five different epochs that are indicated 
by the vertical lines in the relevant panels of Fig.~\ref{fig:images_size_shape_k0},
and span times before, during and after the time when the counter-jet becomes visible.

Fig.~\ref{fig:images_k02} also shows the location of the central source (thin red plus sign), 
and the results of a fit to an elliptical Gaussian, where the best fit ellipse is shown in magenta 
and its center is indicated by a thick magenta plus sign.  The motivation for such a fit is that when 
the image is only marginally resolved (i.e. when its angular size is comparable or slightly smaller 
than the instrumental beam size) one usually performs a fit to the visibility data of a predetermined 
functional form such as a circular or an elliptical data, depending no the quality of the data 
\citep[e.g.][]{Taylor05,TG06,Pihlstrom07,Mesler12}. Because of the reflection symmetry, 
$\tilde{y}\to-\tilde{y}$, the center of the ellipse is along the $\tilde{x}$-axis, at 
$(\tilde{x},\tilde{y})=(\tilde{x}_{\rm el},0)$, and its semi-major/minor axes are along the 
$\tilde{x}$ and $\tilde{y}$ axes (with lengths or standard deviations $\sigma_x$ and $\sigma_y$, respectively). 
The model surface brightness that is fit at each observed time is hence
$I_\nu\propto\exp\left[-(\tilde{x}-\tilde{x}_{\rm el})^2/2\sigma_x^2-\tilde{y}^2/2\sigma_y^2\right]$.

Finally, Fig.~\ref{fig:images_k02} also shows (by a white X sign) the flux centroid's location
on the plane of the sky, which is defined as
\begin{equation}
\tilde{\bm r}_{\rm fc}=(\tilde{x}_{\rm fc},\tilde{y}_{\rm fc}) 
=\frac{\int\,dF_\nu\,(\tilde{x},\tilde{y})}{F_\nu}=\frac{\int\,dF_\nu\,\tilde{\bm r}}{\int\,dF_\nu}\;,
\end{equation}
where $dF_\nu=I_\nu d\Omega=I_{\nu}d_A^{-2}dS_\perp\propto{}I_{\nu}d\tilde{x}d\tilde{y}$.
In our case $\tilde{y}_{\rm fc}=0$ because of the reflection symmetry, $\tilde{y}\to-\tilde{y}$, 
so that the flux centroid's location is $\tilde{\bm r}_{\rm fc}=(\tilde{x}_{\rm fc},0)$ and 
fully specified by its $\tilde{x}$ coordinate, $\tilde{x}_{\rm fc}$.  

Fig.~\ref{fig:images_size_shape_k0} shows the evolution of $\tilde{x}_{\rm fc}$ and of the best fit 
parameters for a fit of the surface brightness (or specific intensity $I_\nu$) of the image to an elliptical Gaussian.
For $\theta_{\rm obs}=0$, $\tilde{x}_{\rm fc}=\tilde{x}_{\rm el}=0$ where the displayed  
$\tilde{x}_{\rm fc}$-values show the numerical accuracy, and are a few decades below $\sigma_x=\sigma_y$.
The fit to an elliptical Gaussian is more reasonable either at early times before the counter-jet 
becomes visible or shortly after it becomes visible and dominates the observed flux. 
Before the counter-jet becomes visible the image is dominated by the main jet that points closer to us, 
and the best fit elliptical Gaussian is centered ($\tilde{x}_{\rm el}$) near the projection of 
the front of this jet onto the plane of the sky (as is the flux centroid, $\tilde{x}_{\rm fc}$), 
while its semi-major axis is perpendicular to the plane containing the jet axis 
and our line of sight (i.e. the $\tilde{y}$ axis; $\sigma_x>\sigma_y$).

Around the time when the counter-jet becomes visible the fluxes from the main jet and 
counter-jet become comparable, corresponding to two rather compact bright regions in the image 
that are separated by an angular distance significantly larger than their own angular size.
At this stage the fit to an elliptical Gaussian becomes quite poor (an alternative fit to two compact 
sources may provide a better fit), and the best fit corresponds to an ellipse that 
is highly elongated along the $\tilde{y}$ axis ($\sigma_y\gg\sigma_x$), whose major axis $2\sigma_y$ 
roughly corresponds to the projected angular separation between the heads of the two jets. 
The counter-jet is more compact and circular at this stage while the jet pointing closer to us
shows a bow-shock like morphology with a somewhat larger angular size. 
At slightly later times when the counter jet dominates the observed flux, the fit to an 
elliptical Gaussian improves, and it is centered around the projected location of the counter-jet's head 
(as is the flux centroid, $\tilde{x}_{\rm fc}$), and becomes more circular ($\sigma_y\approx\sigma_x$).

Comparing the images for $k=0$ and $k=2$ corresponding to the same $\theta_{\rm obs}$ 
and a similar flux ratio between the main jet and counter-jet, it appears that the best fit ellipse has a smaller
axis ratio for $k=2$ compared to $k=0$, corresponding to a somewhat less elongated and rounder image.
This trend is consistent with the images for the spherical self-similar relativistic phase in which the effective 
width of the emitting shell of shocked external medium behind the afterglow shock increases with $k$
\citep{BM76,DeColle12a}, resulting in a more uniform and less limb-brightened image \citep{GL01,Granot08}. 

The relatively rapid transition between the flux being dominated by the main jet and the counter-jet
results in a rather fast motion of the flux centroid $\tilde{x}_{\rm fc}$, as can clearly be seen in 
Fig.~\ref{fig:images_size_shape_k0}, especially in the bottom panels. The maximal
displacement of the flux centroid from the projected location of the central source, 
$\tilde{x}_{\rm max} = \max(|\tilde{x}_{\rm fc}|)$, or the flux centroid's total motion, 
$\Delta\tilde{x}_{\rm fc}$, are expected to be of the order of the jet's (core) 
non-relativistic transition radius, $R_{\rm NR}$, for large viewing angles $\theta_{\rm obs}\approx1$.
It decreases for smaller viewing angles due to the projection effect, such that
\begin{equation}
\tilde{x}_{\rm max}(\theta_{\rm obs}<1)\approx R_{\rm NR}\sin\theta_{\rm obs}\ .
\end{equation}
For for the largest viewing angles, $\theta_{\rm obs}\approx\pi/2$ ($\theta_{\rm obs}=\pi/2$), 
$\Delta\tilde{x}_{\rm fc}$ decreases (vanishes) since in that case the two jets have rather similar
(equal) fluxes and projected displacements around the non-relativistic transition time, 
which causes the flux centroid to be closer to (exactly at) the projected location of the central source.

A more delicate question is how to best estimate $R_{\rm NR}$ \citep[e.g.][]{GL03,GR-RL05,Wygoda11,GP12,DeColle12b}. 
Assuming the jet spreads sideways exponentially once $\Gamma<\theta_0^{-1}$ at $R>R_j$ leads to
\begin{eqnarray}\label{eq:R_NR1}
R_{\rm NR,1}&\approx& (1-\ln\theta_0)R_j\ ,
\\
R_j &=& \fracsb{(3-k)E_{\rm jet}}{2\pi Ac^2}^{1/(3-k)}
= 2^{1/(3-k)}R_{\rm S}(E_{\rm jet}) 
\\ \nonumber
&=&
\left\{\begin{matrix}8.59\times 10^{17}E_{\rm jet,51.3}^{1/3}n_0^{-1/3}\;{\rm cm} 
& (k=0)\ , \cr & \cr
7.06\times 10^{17}E_{\rm jet,51.3} A_*^{-1}\;{\rm cm}  & (k=2)\ , \end{matrix}\right.
\end{eqnarray}
where $R_j$ is the jet break radius, $E_{\rm jet,51.3}=E_{\rm jet}/(2\times10^{51}\;{\rm erg})$ 
and $R_{\rm S}(E_{\rm jet})$ is the Sedov radius corresponding to the jet's true energy.
If, on the other hand, one neglects the jet's lateral spreading (which numerical simulations suggest to be 
modest for $\theta_0\gtrsim0.1-0.2$) and assumes it continues to evolve as if it were part of a spherical 
flow even after the jet break time, until it becomes non-relativistic then $R_{\rm NR}=R_{\rm S}(E_{\rm k,iso})$
corresponds to to the Sedov radius for the jet's isotropic equivalent energy,
\begin{equation}\label{eq:R_NR2}
R_{\rm NR,2}=\left[\frac{(3-k)E_{\rm k,iso}}{4\pi Ac^2}\right]^\frac{1}{3-k} =
\left\{\begin{matrix}2.51\times 10^{18}E_{53}^{1/3}n_0^{-1/3}\;{\rm cm} 
& (k=0)\ , \cr & \cr
1.77\times 10^{19}E_{53} A_*^{-1}\;{\rm cm}  & (k=2)\ . \end{matrix}\right.
\end{equation}
Judging from the jet's dynamics in hydrodynamic simulations \citep[see, e.g., Figs. 4 and 5 of][]{DeColle12b},
and estimating $R_{\rm NR}$ by the jet's radius when its energy weighted mean proper velocity 
$u=\Gamma\beta$ equals unity, it appear to be closer to $R_{\rm NR,1}$ than to $R_{\rm NR,2}$.
From our calculated $\tilde{x}_{\rm max} = \max(|\tilde{x}_{\rm fc}|)$ for $k=0$ we infer
$\tilde{x}_{\rm max}/R_{\rm NR,1}\sin\theta_{\rm obs} = 1.12$ and 1.01 while
$\tilde{x}_{\rm max}/R_{\rm NR,2}\sin\theta_{\rm obs} = 1.00$ and 0.90, 
for $\theta_{\rm obs}=0.4$ and 0.8, respectively, showing a similarly good agreement.
However, for $k=2$ we obtain $\tilde{x}_{\rm max}/R_{\rm NR,1}\sin\theta_{\rm obs} = 2.47$ and 1.99 while
$\tilde{x}_{\rm max}/R_{\rm NR,2}\sin\theta_{\rm obs} = 0.099$ and 0.079, 
for $\theta_{\rm obs}=0.4$ and 0.8, respectively, implying a poorer agreement, and a better 
match for $R_{\rm NR,1}$. A better agreement, good to $\sim10\%$, is obtained when using
\begin{equation}\label{eq:R_NR_approx}
R_{\rm NR}=R_{\rm NR,1}^f R_{\rm NR,2}^{1-f}\quad {\rm with}\quad f\approx0.75\ .
\end{equation}
Note the stronger dependence of $\Delta\tilde{x}_{\rm fc}\sim\tilde{x}_{\rm max}\propto R_{\rm NR}$ 
(and the corresponding angular scale that is discussed next)
on $E_{\rm k,iso}$ or $E_{\rm jet}$ and on the external density normalization $A$ for larger $k$-values.
 
The angular size of the image around the time of the peak in the lightcurve for a given $\theta_{\rm obs}$ 
also scales as $R_{\rm NR}$, and becomes comparable to $\tilde{x}_{\rm max}\sim\Delta\tilde{x}_{\rm fc}$ 
around the time when the counter jet becomes visible and its flux becomes comparable to that of the main jet.  
The corresponding typical angular scale assuming a relatively low redshift source at a distance of 
$D = 100\;D_{100}\;$Mpc, which may potentially be resolved is \citep[e.g.][]{GL03},  is
\begin{eqnarray}\nonumber
\theta_{\rm NR} = \frac{R_{\rm NR}}{D}=
\left\{\begin{matrix}1.54\,g_{0.2}\theta_{0.2}^{-1/6}E_{\rm jet,51.3}^{1/3}n_0^{-1/3}D_{100}^{-1}\;{\rm mas} 
& (k=0)\ , \cr & \cr
1.67\,g_{0.2}\theta_{0.2}^{-1/2}E_{\rm jet,51.3}A_*^{-1}D_{100}^{-1}\;{\rm mas}  & (k=2)\ , \end{matrix}\right.
\\ \label{eq:theta_NR}
\hfill=
\left\{\begin{matrix}1.54\,g_{0.2}\theta_{0.2}^{1/2}E_{53}^{1/3}n_0^{-1/3}D_{100}^{-1}\;{\rm mas} 
& (k=0)\ , \cr & \cr
1.67\,g_{0.2}\theta_{0.2}^{3/2}E_{53}A_*^{-1}D_{100}^{-1}\;{\rm mas}  & (k=2)\ , \end{matrix}\right.
\quad
\end{eqnarray}
where $g_{0.2}=[(1-\ln\theta_0)/(1-\ln0.2)]^{0.75}$, $\theta_{0.2}=\theta_0/0.2$, and we have
used Eq.~(\ref{eq:R_NR_approx}). For comparison, the Very Long Baseline Array (VLBA) has an angular 
resolution of $\sim170\;\mu$as at 43~GHz, and may potentially resolve the jet around the time of the peak 
in the lightcurve for binary mergers that are detectable in gravitational waves by advanced LIGO/VIRGO.

\section{Scaling with Model Parameters and Degeneracies}
\label{sec:scaling}

Inferring all of the model parameters from detailed fits to afterglow data is usually a challenging task, 
even when elaborate observations are available, due to the rather large number of model free parameters, 
and degeneracies between them. Nonetheless, for well monitored afterglows some of the keys model 
parameters can be inferred reasonably well, such a the electron power-law index $p$, which can be 
derived from the spectral slope in PLSs G or H. Ideally, for an on-axis observer the temporal decay index 
in the same PLS (for a spherical flow or before the jet break time) could then help determine the 
external density power-law index, $k$. Then, the parameters $\epsilon_e$, $\epsilon_B$, $E_{k,iso}$, 
and the external density normalization $A$ can be determined by the flux normalization $F_{\nu,max}$ 
and three break frequencies $\nu_{sa}$, $\nu_m$ and $\nu_c$ \citep[e.g.][]{WG99,SE01,GR-RL05}, 
up to the degeneracy pointed out by \citet{EW05}. The latter degeneracy arises from the uncertainty on 
the fraction $\xi_e$ of the post-shock accelerated electrons that take part in the power-law energy 
distribution that emits the synchrotron radiation we observe. For a jet there are additional 
free parameters, namely our viewing angle $\theta_{\rm obs}$ relative to the jet's symmetry axis, and
parameters that describe its initial angular structure (its initial half-opening angle $\theta_0$ for a 
top-hat jet, and usually more parameters for  other jet structures).

In practice, even in some very well monitored afterglows and when we have good reason to expect $k=0$, 
such as for the short GRB$\,$170817A$\,$/$\,$GW$\,$170817, a lot of degeneracy still remains even after 
a very detailed fit to the afterglow lightcurves at all observed frequencies. For this reason imaging becomes 
a very important diagnostic tool that may potentially help to break such a degeneracy \citep[e.g.,][]{GG18,NP18,Nakar18}. 

Fitting afterglow data to the results of numerical calculations based on hydrodynamical simulations of the 
GRB jet during the afterglow phase becomes much more efficient numerically when taking advantage of
the relevant scaling relations \citep[e.g.][]{Granot12,vEvHM12,vanEerten12a}. This scaling ultimately 
arises from the freedom in the choice of the three basic physical units (of mass, length and time) when 
applying the results of a numerical simulation to the relevant physical system \citep{Granot12}. 
Relativistic hydrodynamic (or magneto-hydrodynamic; MHD) simulations must preserve the value of 
the speed of light in vacuum $c$ (a universal dimensional constant), requiring the scaling factors of length 
and time to be equal, thus leaving two free parameters for rescaling simulation results: $\alpha = t'/t=l'/l$
and $\zeta = m'/m$ when rescaling to primed units and quantities \citep[see][for details]{Granot12}. 
Instead of using the scaling factors of the basic physical units, one can conveniently use those for 
useful physical quantities such as the energy $\kappa = E'/E = m'/m = \zeta$ and proper rest mass density
$\lambda = \rho'/\rho = \zeta/\alpha^3$ \citep[e.g.][]{vEvHM12,vanEerten12a}. 

In our case it may be more convenient to rescale the external density normalization factor 
$a = A'_k/A_k = \zeta/\alpha^{3-k}=\lambda^{(3-k)/3}\kappa^{k/3}$ and energy $\zeta=\kappa=E'/E$. 
In this case length and time scale by a factor $\alpha = (\zeta/a)^{1/(3-k)}=(\kappa/\lambda)^{1/3}$. 
This can be seen in Eqs.~(\ref{eq:R_NR1})-(\ref{eq:R_NR2}) where the critical radii and in particular 
$R_{\rm NR}$ scale as $(E/A)^{1/(3-k)}$. Since $R'(t'=\alpha t) = \alpha R(t)$ one can conveniently 
normalize the lengths and times by $R_{\rm NR}$ and $t_{\rm NR}=R_{\rm NR}/c$, respectively, 
$\bar{t}=t/t_{\rm NR}$ and $\bar{l}=l/R_{\rm NR}$. In these normalized units the size and shape 
of the image at any given observed time (as well as the normalized surface 
brightness distribution within the image at any given spectral PLS), and in particular the ones that are
shown in Fig.~\ref{fig:images_k02} are valid for any rescaling of the energy ($\zeta$) and the
external density normalization factor ($a$), which only affect $R_{\rm NR}=ct_{\rm NR}\propto(E/A)^{1/(3-k)}$.
Therefore, measurements of the image size can help constrain $E/A$.

While the scaling factor $\alpha$ of length and time depends only on the ratio of the scaling factors, 
$\zeta/a$, the scaling of the flux density $F_\nu$ within each spectral PLS depends on each of the 
scaling factors separately, where the dependence changes between different PLSs
\citep[for the explicit scalings see][]{Granot12,vanEerten12a}. Note that within each PLS the usual 
dependence on the shock microphysics parameters ($\epsilon_e$, $\epsilon_B$, $\xi_e$, $p$) remains valid
\citep[e.g.][]{GS02,vanEerten12a,Granot12}. For any rescaling by factors $(\zeta,a)$, 
within each PLS $t$ scales by a factor $\alpha = (\zeta/a)^{1/(3-k)}$ while $F_\nu$ scales by another 
PLS-dependent factor. In a log-log plot of $F_\nu(t)$ this corresponds to horizontal and vertical shifts
of the lightcurve, along the time and flux density axes, respectively, while its shape does not change. 
The lightcurve shape depends on the dynamics, namely on the external density power-law index, $k$, 
and the jet angular structure, which may make it possible to constrain $k$, even when some degeneracy 
remains in the other model parameters. 

The scaling of $F_\nu$ implies that the mean surface brightness within the image, $\mean{I_\nu}$, must 
also scale correspondingly ($\mean{I_\nu}\propto F_\nu/S_\perp$ where $S_\perp\propto l^2$ is the area 
of the image on the plane of the sky, such that $S'_\perp/S_\perp = \alpha^2$), and has the same frequency
dependence as $F_\nu$ within any given PLS. However, within each PLS the normalized surface brightness, 
$I_\nu/\mean{I_\nu}$, as a function of the normalized location within the image at any given normalized time, 
$[\tilde{x}(\,\bar{t}\,),\,\tilde{y}(\,\bar{t}\,)]/R_{\rm NR}$,  remains invariant under any rescaling by factors 
$(\zeta,a)$. All of the scalings mentioned above make our results applicable to a wide range of parameter space.

\section{Discussion}
\label{sec:dis}
 
Off-axis lightcurves from 2D relativistic hydrodynamic simulations have been presented for different 
viewing angles $\theta_{\rm obs}$ with respect to the symmetry axis of a jet propagating into a 
power-law external density profile, $\rho_{\rm ext}\propto{}R^{-k}$ for $k=0,\,1,\,1.5,\,2$, ranging 
from a uniform ISM-like medium ($k=0$) that is expected for short GRBs, to a stratified (steady) 
wind-like medium ($k=2$) that may be expected from the massive star progenitors of long GRBs. 
The lightcurves were calculated in the radio, optical and X-ray, as such orphan afterglows may be 
detected in upcoming surveys covering different parts of the electromagnetic spectrum. 
It was found that for off-axis observers ($\theta_{\rm obs}>\theta_0$) a larger $k$ results in a
shallower the rise to the peak of the lightcurve with the flatter and wider the peak, leading to 
a much less pronounced bump in the afterglow lightcurve when the counter-jet becomes visible 
that hard to clearly observe for $k=2$.

This may potentially partly explain the lack of a clear counter-jet induced bump in the late afterglow
lightcurves of long GRBs, for which $1\lesssim k\lesssim 2$ may be expected.
For the longest GRB afterglow monitored in the radio, GRB$\;$030329, it is not clear how well 
such an explanation for the lack of a clear flattening or rebrightening \citep[e.g.][]{Pihlstrom07,Mesler12} 
might work, since in that case detailed afterglow modeling favors a uniform
external density \citep[$k = 0$;][]{vdH08}. It is worth noting, however, that for nearby NS-NS or NS-BH
mergers that are detected in gravitational waves and are accompanied by long-lived afterglow emission,
a uniform external medium is expected ($k=0$), which may help in detecting a late time flattening or 
rebrightening in the lightcurve corresponding to the contribution from the counter-jet.
It would be useful to search for such a signal, which may help probe the structure of the outflow
from such events, and the symmetry between the main jet and counter jet, and/or the external density
that they are expanding into.

The corresponding afterglow images were also calculated in the radio, as that is where the best 
angular resolution is currently available, using very large baseline interferometry (VLBI; see the 
discussion around Eq.~(\ref{eq:theta_NR})). In particular, the observed size and shape of the radio 
afterglow image were calculated along with the motion of its flux centroid, which may be measured 
even in some cases when the image itself is not resolved. Fits of the image to an elliptical Gaussian 
were also performed, since they are often done by observers when the image is only marginally resolved, 
and their detailed properties were discussed.

These detailed properties of the afterglow lightcurves and image may help to clearly distinguish 
orphan GRB afterglows from other types of transients in upcoming surveys, which may otherwise
prove to be very challenging. In particular, this may help identify relativistic jets that are pointed 
away from us, either in nearby supernovae Ib/c (some of which have been associated with long 
duration GRBs)  for which $1\lesssim{}k\lesssim 2$ may be expected, or in nearby binary neutron star 
mergers that are detected through their gravitational wave signal, and may also produce short duration 
GRBs at least for some viewing angles (as in the case of GW$\,$170827/GRB$\,$170817A).
It is most promising to detect or angularly resolve such transients near the time of the peak in their lightcurve.
which for large viewing angles corresponds to a Lorentz factor $\Gamma\lesssim\;$a few.
Therefore, most of the results in this work are applicable also for moderately relativistic jets with a modest 
initial Lorentz factor of $\Gamma_0\gtrsim\;$a few, which may be intrinsically much more common than
ultra-relativistic jets ($\Gamma_0\gg1$ or $\Gamma_0\gtrsim100$ that are often inferred for GRBs).

\section*{Acknowledgements}

J.G. is supported by the Israeli Science Foundation under grant No. 719/14.
FDC aknowledges support from the UNAM-PAPIIT grant IN117917.
ER-R is supported in part by  David and Lucile Packard Foundation and
the Niels Bohr Professorship from the DNRF.
We acknowledge the support from the Miztli-UNAM supercomputer (project LANCAD-UNAM-DGTIC-281) 
in which the simulations were performed.




\bsp	
\label{lastpage}
\end{document}